# Galaxies: kinematics as a proof of the existence of a universal field of minimum acceleration

A Alfonso-Faus

E.U.I.T. Aeronáutica, Plaza Cardenal Cisneros, s/n, 28040 Madrid, Spain

E-mail: aalfonsofaus@yahoo.es

**Abstract.** Taking into account only luminous objects, the kinematics of clusters of galaxies, galaxies and their interior, require a much higher mass than the luminous one to explain the observations. This situation has provoked more than 30 years of intense research and has stimulated hypothesis like the dark matter. Also new mechanical theories, different from Newton (like the Modified Newtonian Dynamics), have been proposed. We here present an alternative: theoretical and observational data strongly suggest the existence of a universal field of minimum acceleration of the order of $c/t$, $c$ the speed of light and $t$ the age of the Universe. We keep our reasoning within the present state of the art of Quantum Mechanics, Relativity and Newtonian Mechanics. With this approach the kinematics of the luminous objects are explained without any additional assumption. At the same time the sizes of all structures in the Universe, from protons up to the Universe, are explained as due to the constant action of this field.

## 1. Introduction: observational data

Observing gravitational effects in luminous objects, like stars, gas clouds, globular clusters, even entire galaxies, brought up the possibility of existence of dark matter Jungman et al., (1996) many years ago. The hypothesis was needed to explain the lack of enough mass in these systems to account for the observed gravitational effects. We present the evidence from clusters of galaxies and the rotation curves of spiral galaxies.

Measurements of the total masses of some clusters of galaxies, Zwicky (1933) and Smith (1936), showed that they could not bind the galaxies together unless much more mass were present (and non-luminous). Hence the existence of some kind of dark matter was assumed.

Years later, Rubin and Ford (1970), Faber and Gallagher, (1979), Rubin et al. (1980), the rotation curves of spiral galaxies were determined and found to be discrepant with the classical theory: objects far from the centre of the galaxy should have a decreasing speed with distance. The observed values indicated a flat curve, or even an increasing speed with distance. Applying Newton's laws one would fit observations postulating the existence of dark matter with a total mass for the galaxy proportional to the radial distance. But we do not "see" this dark matter by any other means. An alternative is to modify Newton's laws accordingly to explain the observations, Milgrom (1983).

In what follows we present theoretical arguments that strongly suggest the existence of a universal acceleration field. It fully explains the kind of data we have given above, without the need of dark matter or any new mechanical theory. It also explains the sizes of astrophysical objects, including the Universe and the fundamental particles.

## 2. Universal field of minimum acceleration

Gravitation permeates all space. At every point in space the gravitational attraction of the rest of the visible Universe is acting. Taking into account the cosmological principle, space homogeneous and isotropic, a point mass has to feel the attraction of the rest of the Universe but totalling a zero resultant force because of the symmetry involved. And therefore the resultant acceleration is also zero due to the isotropic distribution. It has spherical symmetry: the same from any direction. But local anisotropies around a test mass certainly distort this picture and make the isotropic vectors to adapt to the anisotropy. Hence the distribution of the acceleration field acting around the test particle is distorted to a new configuration having a preferred direction pointing towards the centre of mass of the nearby objects. The same has to be for all objects in the halo of galaxies: the anisotropic universal acceleration has to point towards the centre of the galaxy. These objects are subject to the universal acceleration, non isotropic, pointing towards the centre of each galaxy. And for clusters of galaxies we have the same picture. It is a universal effect.

If M is the mass of the visible Universe, R its size of the order of ct, c the speed of light and t the age of the Universe, then the universal acceleration has the following order of magnitude:

$$\frac{GM}{R^2} = \frac{GM}{c^2} \cdot \frac{1}{t^2} \approx \frac{R}{t^2} \approx \frac{c}{t} \qquad (1)$$

Here we have used a numerical "coincidence" pointed out by Weinberg (1972) that in fact comes from the Einstein's cosmological equations. These two equations have terms of the order of $G\rho$, $\rho$ the mass density of the Universe, and other terms of the order of $1/t^2$. Then one has

$$G\rho t^2 \approx 1 \qquad (2)$$

For example, using a mass density for the Universe of about $10^{-28}$ gr/cc and an age of about $1.4 \times 10^{10}$ years one gets 1.3 for the numerical order in (2). Since the mass density is of the order of $M/R^3$ and assuming a Machian Universe of size R = ct, one finally gets

$$\frac{GM}{R^3} \approx \frac{1}{t^2} \approx \frac{c^2}{R^2} \quad i.e. \quad \frac{GM}{c^2} \approx R \approx 10^{28} cm \qquad (3)$$

which is a form of Mach's principle stating that the gravitational potential energy of any mass m i.e. GMm/R, with respect to the mass of the rest of the Universe M, is of the order of its relativistic energy $mc^2$. Then m drops from both sides of this relation and we get the Machian result (3): the gravitational radius of the Universe is of the order of its size.

3. The minimum acceleration from quantum gravity

The minimum quantum of action is Planck's constant ℏ. The maximum interval of time is the age of the Universe t. Then the minimum quantum of energy possible in the Universe is ℏ/t; a quantum (ℏ) cosmology (t) argument. The ground state of the minimum-energy quantum of gravity Alfonso-Faus (2000) has this energy. And its relativistic minimum mass $m_g$ is obtained dividing this energy by $c^2$ i.e.

$$m_g = \frac{\hbar}{c^2 t} \approx 2 \times 10^{-66} \, grams \tag{4}$$

This is the relativistic mass of the quantum of gravity that we have proposed elsewhere Alfonso-Faus (2000). We can now introduce the concept of a minimum force in the Universe. Since we have the minimum energy in (4) as ℏ/t, dividing by the maximum length ct we get the minimum force in the Universe $F_{min}$ as

$$F_{min} = \frac{\hbar}{ct^2} = m_g \cdot a_{min} \tag{5}$$

Here we have used the obvious condition that the minimum force must be equal to the product of the minimum mass in (4) times the minimum acceleration that we are looking for Alfonso-Faus (2000). Then we get for this:

$$a_{min} = \frac{F_{min}}{m_g} = \frac{c}{t} \tag{6}$$

which is the same as the acceleration found in (1). This is really a minimum acceleration quantized field which comes from quantum mechanical arguments in conjunction with a Machian approach: the gravitational action of the whole visible Universe. And all space is permeated by this field. It would be a homogeneous and isotropic field in the same sense as the cosmological principle states. But local distribution of masses distorts the field that gets anisotropic. Then the final effective acceleration $a_e$ acting upon a test mass, at distance r from the centre of gravity of a mass M, will be lower by a factor $\alpha$

$$a_e = \alpha \frac{c}{t}, \qquad 0 \leq \alpha \leq 1 \tag{7}$$

where the coefficient $\alpha$ takes into account the resultant anisotropy upon the field at distance r from M. This coefficient must be a function of r and M. We will see that all the dark matter evidence can be substituted by the action of this field.

## 4. Rotation curves in spiral galaxies

Newton's law for luminous objects in near circular orbits of size r at speed v, orbiting the centre of a galaxy of total luminous mass $M_g$ gives the following result at the halo (total attracting luminous mass constant, $M_g$):

$$v^2 = \frac{GM_g}{r} \tag{8}$$

The observed rotation curves imply that the speed v, instead of decreasing with the distance r as in (8), is constant or even increases slowly when far from the central luminous object, Drees et al. (2007). Here comes the strongest "evidence" for dark matter. If $M_g$ in (8) instead of being constant is supposed to increase with r, then we can fit all observations by introducing an ad hoc hypothesis: a dark halo with the appropriate non-luminous mass distribution of dark matter to fit observations. At the distances involved the acceleration due to the gravitational attraction of the constant mass M (no dark matter) is becoming much lower than the minimum $\alpha(c/t)$ that we have found in the previous sections. Hence far enough from M, like in the halos of galaxies, the minimum prevails and we have a constant acceleration $\alpha(c/t)$ directed towards the centre of the galaxies. Again, for near circular orbits we get the centripetal acceleration:

$$\frac{v^2}{r} = a_e = \alpha \frac{c}{t} \tag{9}$$

This gives an increasing speed with r, as observed in many cases Drees et al. (2007). However, one expects that the anisotropy responsible for the acceleration in (9) has to decrease with distance r (it tends to zero as r increases) and has to increase with $M_g$. One interesting approach is to assume an inverse linear dependence from the value of the gravitational size $r_G$ onwards. This gravitational size is given below, in section 5, by the expression (12). Then one has for $r \geq r_G$

$$\frac{v^2}{r} = a_e = \frac{r_G}{r}\frac{c}{t} \tag{10}$$

Then the rotation curves tend to a constant speed $v_0$ given by

$$v_0^2 = \frac{c}{t}r_G = \frac{GM_g}{r_G} \tag{11}$$

Using typical values for $M_g = 2.4 \times 10^{44}$ grams and $r_G = 4 \times 10^{22}$ cm, from relation (12), one gets $v_0$ = 200 Km/sec which is a typical constant speed observed for the rotation curves Drees et al. (2007).

**5. Formation of structures from clusters of galaxies to fundamental particles**

The universal minimum acceleration field can explain the gravitational formation of all structures in the Universe, from the clusters of galaxies to the fundamental particles. If one takes into account Newton's law (8) when $v^2$ is substituted into the acceleration strength by the universal acceleration field (9) (which implies the end of the gravitational mass extension M in space and the beginning of free space) one gets for the gravitational size of objects $r_G$, following (11)

$$r_G^2 \approx \frac{Gt}{c}M \approx \frac{GM}{c^2}ct \approx r_g R \tag{12}$$

where $r_g$ is the gravitational radius of the mass M and R the gravitational radius of the Universe, about its size. It is a geometric inversion law. This gives a gravitational size $r_G$ for all structures in the Universe. If one introduces the actual size and mass of structures like the whole Universe, clusters of galaxies, galaxies, globular clusters, nebulae, and even the fundamental particles, one can check that the relation (12) is very well satisfied for all these structures (see Table 1). The conclusion is that the Universal field of minimum acceleration has had a very strong and a very deep influence in forming all structures in the Universe. In fact it is the reason for them to exist.

**6. Conclusions**

Theoretical and observational data strongly suggest the existence of a universal field of minimum acceleration of order c/t. No additional assumptions, like the dark matter hypothesis or any modification of Newton's laws, are necessary to explain what is observed. The proposed existence of the universal field c/t is reinforced by the explanation of the sizes of all structures in the Universe. They are due to the action of this field, which permeates all space and has a Machian origin. It comes from the presence of all visible matter in the Universe.

TABLE 1  Astrophysical Objects

| Object | Mass | Gravitational radius $r_g$ | $r_g.R$ | Size $r = (r_g.R)^{1/2}$ |
|---|---|---|---|---|
| Universe | $10^{56}$ grams | $R = 10^{28}$ cms | $10^{56}$ cm$^2$ | $10^{28}$ cms |
| Cluster of galaxies | $10^{47}$ grams | $10^{19}$ cms | $10^{47}$ cm$^2$ | $10^{24}$ cms |
| Galaxies | $2 \times 10^{44}$ grams | $2 \times 10^{16}$ cms | $2 \times 10^{44}$ cm$^2$ | $1.5 \times 10^{22}$ cms |
| Globular clusters | $2 \times 10^{39}$ grams | $2 \times 10^{11}$ cms | $2 \times 10^{39}$ cm$^2$ | $5 \times 10^{19}$ cms |
| Nebulae | $2 \times 10^{38}$ grams | $2 \times 10^{10}$ cms | $2 \times 10^{38}$ cm$^2$ | $1.5 \times 10^{19}$ cms |
| Protostar | $2 \times 10^{33}$ grams | $2 \times 10^{5}$ cms | $2 \times 10^{33}$ cm$^2$ | $4 \times 10^{16}$ cms |
| Protons | $1.7 \times 10^{-24}$ grams | $1.7 \times 10^{-52}$ cms | $1.7 \times 10^{-24}$ cm$^2$ | $10^{-12}$ cms |

From $r = 10^{-12}$ cms (protons) up to $r = 10^{28}$ cms (Universe) the geometric inversion implied by the relation $r_G^2 = r_g . R$ is remarkably well satisfied by the Universe, cluster of galaxies, galaxies, globular clusters, nebulae, protostar systems and protons.